\documentclass[useAMS,usenatbib,amsmath,twocolumn]{mn2e}
\usepackage{graphicx}
\usepackage{color}
\usepackage{amssymb}
\usepackage{amsmath}
\input{colordvi.tex}

\title{21cm-line bispectrum as a method to probe Cosmic Dawn and Epoch of Reionization}
\author[H.Shimabukuro et al.]
  {Hayato Shimabukuro$^1$$^{,2}$,
  Shintaro Yoshiura$^2$, Keitaro Takahashi $^2$, Shuichiro Yokoyama$^3$
  \newauthor 
   and Kiyotomo Ichiki $^1$ \\
  $^1$Department of Physics, Graduate School of Science, 
Nagoya University, Aichi, 464-8602, Japan\\
  $^2$Department of Physics, Kumamoto University, Kumamoto, Japan\\
  $^3$Department of Physics, Rikkyo University, Tokyo, Japan}

\pagerange{\pageref{firstpage}--\pageref{lastpage}} \pubyear{2002}

\def\LaTeX{L\kern-.36em\raise.3ex\hbox{a}\kern-.15em
    T\kern-.1667em\lower.7ex\hbox{E}\kern-.125emX}

\begin{document}

\label{firstpage}

\maketitle

\begin{abstract}
Redshifted 21cm signal is a promising tool to investigate the state of intergalactic medium (IGM) in the Cosmic Dawn (CD) and Epoch of Reionization(EoR). In our previous work \cite{2015MNRAS.451.4986S}, we studied the variance and skewness of the 21cm fluctuations to give a clear interpretation of the 21cm power spectrum and found that skewness is a good indicator of the epoch when X-ray heating becomes effective. Thus, the non-Gaussian feature of the spatial distribution of the 21cm signal is expected to be useful to investigate the astrophysical effects in the CD and EoR. In this paper, in order to investigate such a non-Gaussian feature in more detail, we focus on the bispectrum of the 21cm signal. It is expected that the 21cm brightness temperature bispectrum is produced by non-gaussianity due to the various astrophysical effects such as the Wouthysen-Field (WF) effect, X-ray heating and reionization. We study the various properties of 21cm bispectrum such as scale dependence, shape dependence and redshift evolution. And also we study the contribution from each component of 21cm bispectrum. We find that the contribution from each component has characteristic scale-dependent feature.  In particular, we find that the bulk of the 21cm bispectrum at $z$~=20 comes from
the matter fluctuations, while in other epochs it is mainly determined by the
spin and/or neutral fraction fluctuations and it is expected that we could obtain more detailed information on the IGM in the CD and EoR by using the 21cm bispectrum in the future experiments, combined with the power spectrum and skewness.
\end{abstract}

RUP-15-16

\section{Introduction}

After the recombination of protons and electrons in the primordial plasma, the ``dark age'' begins and continues until the first object is formed through the gravitational instability in dark matter halos.  The universe dawns by the lights from the first stars and astrophysical processes start to play important roles in the intergalactic medium (IGM). \cite{yos,Fialkov:2013uwm,Visbal:2012aw,2011A&A...527A..93S}. Among others, Wouthuysen Field (WF) effect due to the Lyman alpha (Ly-$\alpha$) radiation emitted from the primordial stars is expected to be the first important process for IGM, which couples the spin temperature of neutral hydrogen to the color temperature of Ly-$\alpha$ radiation \cite{wou}. In the typical environments of IGM, color temperature of Ly-$\alpha$ radiation field couples with kinetic temperature of IGM because of the large Ly-$\alpha$ scattering rate. Therefore, the spin temperature also couples to kinetic temperature.  The second astrophysical process which is related to the thermal history of IGM after the WF effect is X-ray heating \cite{Pritchard:2006sq}. The X-rays are thought to be emitted from such sources as galaxies and X-ray binaries \cite{Fialkov:2014kta}, and kinetic temperature of IGM increases dramatically \cite{Mesinger:2012ys, Christian:2013gma}. Finally, the ``Epoch of Reionization (EoR)'' \cite{Fan:2006dp} follows these processes when the gas density of IGM decreases enough so that recombination process becomes inefficient. This is the epoch when neutral hydrogens start to be ionized by UV radiation photons from early galaxies \cite{Loeb:2000fc}.

The redshifted 21cm line from neutral hydrogen due to the hyperfine transition is suitable for studying thermal and ionized states of IGM as well as the first objects in the dark age and EoR \cite{fur, 2012RPPh...75h6901P, 2014PhRvD..90h3003S}. One of the statistical methods to subtract the information about the physical state of IGM at those epochs from 21cm line is the power spectrum analysis of brightness temperature \cite{fur, Pritchard:2006sq,2008ApJ...689....1S,Baek:2010cm,Mesinger:2013nua,2014ApJ...782...66P}.  On-going radio interferometers, such as Low Frequency Array (LOFAR) \cite{Rottgering:2003jh}, Murchison Wide Field Array (MWA) \cite{Tingay:2012ps} and Probing the Epoch of Reionization (PAPER) \cite{Pober:2014aca}, have started observation and serve as a ``prototype'' of future high-sensitivity experiments. Although the sensitivities of the on-going experiments are inefficient to form images of the distribution of neutral hydrogen using the redshifted 21cm line from the cosmic dawn and EoR, the power spectrum of the 21cm signal would be detected \cite{Mesinger:2013nua}. In particular, it is expected that the 21cm power spectrum for each redshift ($z = 10 \sim 30$) can be measured by the Square Kilometre Array (SKA) with high accuracy \cite{Carilli:2014vha}.

In our previous work \cite{2015MNRAS.451.4986S}, we gave an interpretation to the time evolution of the 21cm power spectrum and we find that the size of skewness is sensitive to the epoch when X-ray heating becomes effective. Other work also reports the impact of spin temperature fluctuations on the skewness \cite{2015arXiv150507108W} and there is a work which focuses on the redshift distortion as the indicator of the epoch where X-ray heating is effective\cite{2015PhRvL.114j1303F}. Herein we extend these previous works by considering the the bispectrum to investigate the dependence of the skewness on scales because skewness is an integral of the bispectrum with respect to the wave number (see Appendix A). In a different work \cite{2015MNRAS.451.4785Y} we have already estimated errors from the thermal noise of detectors in estimating the bispectrum and we found that the 21cm bispectrum would be detectable at large scales at $k\le 0.1 {\rm Mpc}^{-1}$ even by the current detectors on, such as, the MWA and PAPER. Furthermore, the 21cm bispectrum would be detectable even at small scales with the SKA \cite{2015MNRAS.451.4785Y}, and therefore the study of the bispectrum is timely and well motivated. Some previous works studied the 21cm bispectrum \cite{Cooray:2004kt,Pillepich:2006fj, Cooray:2008eb, 2015arXiv150604152M,Cooray:2006km}. These works, however, mainly focus on the bispectrum as a measurement of primordial non-gaussianity in matter fluctuations. In our study, we instead focus on non-Gaussianity in 21cm fluctuations induced by astrophysical effects, whose size is expected to be larger than that in matter fluctuations. (For other probes of non-Gaussianity such ans Minkowski functionals, see \cite{2006MNRAS.370.1329G,2008ApJ...675....8L,2014JKAS...47...49H,yoshi2015}.)

\section{Formulation and set up}

\subsection{Formulation for the 21cm bispectrum}

A fundamental quantity of 21cm line is the brightness temperature, which is described as the spin temperature offsetting from CMB temperature, given by (see, e.g, \cite{fur})
\begin{align}
\delta T_{b}(\nu) &= \frac{T_{{\rm S}}-T_{\gamma}}{1+z}(1-e^{-\tau_{\nu_{0}}})  \nonumber \\
                  &\quad \sim 27x_{{\rm H}}(1+\delta_{m})\bigg(\frac{H}{dv_{r}/dr+H}\bigg)\bigg(1-\frac{T_{\gamma}}{T_{{\rm S}}}\bigg) \nonumber \\
                  &\quad \times \bigg(\frac{1+z}{10}\frac{0.15}{\Omega_{m}h^{2}}\bigg)^{1/2}\bigg(\frac{\Omega_{b}h^{2}}{0.023}\bigg) [{\rm mK}].
\label{eq:brightness}
\end{align}
Here, $T_{\rm S}$ and $T_{\gamma}$ respectively represent gas spin temperature and CMB temperature, $\tau_{\nu_{0}}$ is the optical depth at the 21cm rest frame frequency $\nu_{0} = 1420.4~{\rm MHz}$, $x_{\rm H}$ is neutral fraction of the hydrogen gas, $\delta_{m}({\bf x},z) \equiv \rho/\bar{\rho} -1$ is the evolved matter overdensity, $H(z)$ is the Hubble parameter and $dv_{r}/dr$ is the comoving gradient of the gas velocity along the ling of sight.  All quantities are evaluated at redshift $z = \nu_{0}/\nu - 1$.

Let us focus on the spatial distribution of the brightness temperature. The spatial fluctuation of the brightness temperature  can be defined as
\begin{eqnarray}
\delta_{21}({\bf x}) \equiv
\delta T_b({\bf x}) - \langle \delta T_b \rangle 
\end{eqnarray}
where $\langle \delta T_b \rangle$ is the mean brightness temperature obtained from brightness temperature map and $\langle ...\rangle$ expresses the ensemble average. From this definition, we have the power spectrum of $\delta_{21}$ defined as
\begin{equation}
\langle \delta_{21}({\bf k}) \delta_{21}({\bf k^{'}})\rangle
= (2\pi)^3 \delta({\bf k}+{\bf k^{'}}) P_{21}({\bf k}),
\label{eq:ps_def}
\end{equation}
If the statistics of the brightness temperature fluctuations is pure Gaussian, the statistical information of the brightness temperature should be completely characterized by the power spectrum, and in the above expression for the brightness temperature given by Eq. (\ref{eq:brightness}), if both of the spin temperature and the neutral fraction are  completely homogeneous, the statistics of the brightness temperature fluctuations completely follows that of the density fluctuations $\delta_m$. However, in the era of CD and EOR, it is expected that the spin temperature and the neutral fraction should be spatially inhomogeneous and the statistics of the spatial fluctuations of those quantities would be highly non-Gaussian due to the various astrophysical effects. Accordingly, the statistics of the brightness temperature fluctuations would deviate from the pure Gaussian and it should be important to investigate the non-Gaussian feature of the brightness temperature fluctuations. Although such a non-Gaussian feature can be investigated through the skewness of the one-point distribution functions as done in our previous work \cite{2015MNRAS.451.4986S}, the scale-dependent feature has been integrated out in the skewness. On the other hand, the higher order correlation functions in Fourier space such as a bispectrum and a trispectrum characterize the non-Gaussian features and also have the scale-dependent information. Here, in order to see the non-Gaussian feature of the brightness temperature fluctuations $\delta_{21}$, we focus on the bispectrum of $\delta_{21}$ which is given by
\begin{equation}
\langle \delta_{21}({\bf k_{1}}) \delta_{21}({\bf k_{2}}) \delta_{21}({\bf k_{3}})\rangle = (2 \pi)^3 \delta({\bf k_{1}}+{\bf k_{2}}+{\bf k_{3}})B({\bf k_{1}},{\bf k_{2}},{\bf k_{3}}).
\label{eq:bs_def}
\end{equation}
In order to characterize the shape of the bispectrum in $k$-space,
we use an isosceles ansatz which is defined as $k_1 = k_2 = k = \alpha k_3$ ($\alpha \geq 1/2$).
For examples, in case with $\alpha \gg 1$ the shape of  the bispectrum is often called as ``{\it squeezed type}'' or ``{\it local type}'',
in case with $\alpha = 1$ it is called as `` {\it equilateral type}'', and
in case with $\alpha = 1/2$ it is called as ``{\it folded type}''.  Note that we relax the configuration condition because we calculate the bispectrum from the grid point.  We regard the length within the range of $10\%$ of side of the triangle we desire as the that of triangle.

\subsection{Calculation of the 21cm bispectrum}

In this paper, we calculate the bispectrum of the brightness temperature fluctuations (21cm bispectrum) by making use of 21cmFAST \cite{Mesinger:2007pd,2011MNRAS.411..955M}. This code is based on a semi-analytic model of star/galaxy formation and reionization, and makes maps of matter density, velocity, spin temperature, ionized fraction and brightness temperature at the designated redshifts. 

We perform simulations in a $(200 {\rm Mpc})^3$ comoving box with $300^3$ grids, which corresponds to 0.66 c{\rm Mpc} resolution or $\sim$ 12.7(14.1) arcsec at 80 (127) {\rm MHz} (${\it z}$ = 17 (10)) and $1.07 (1.19) {\rm deg}^{2}$ field of view at 80 (127) {\rm MHz} (${\it z}$ = 17 (10)), from $z = 200$ to $z = 8$ adopting the following parameter set, $(\zeta, \zeta_{X}, T_{\rm vir}, R_{\rm mfp}) = (31.5, 10^{56}/M_{\odot}, 10^4~{\rm K}, 30~{\rm Mpc})$. Here, $\zeta$ is the ionizing efficiency, $\zeta_{X}$ is the number of X-ray photons emitted by source per solar mass, $T_{\rm vir}$ is the minimum virial temperature of halos which produce ionizing photons, and $R_{\rm mfp}$ is the mean free path of ionizing photons through the IGM. In our calculation, we also ignore, for simplicity, the gradient of peculiar velocity whose contribution to the brightness temperature is relatively small (a few \%) \cite{Ghara:2014yfa}. We perform 10 realizations of simulations with different initial condition of density fluctuations and obtain brightness temperature maps. Then we evaluate the average bispectrum as
\begin{eqnarray}
abs[\overline{B(k)}] &=& \frac{1}{N}\sum_{i=1}^{N}abs[B(k)]_{i} \nonumber \\
&=&\frac{1}{N}\sum^{N}_{i}({\rm Re}[B(k)]^{2}+{\rm Im}[B(k)]^{2})_{i=1}^{1/2}.
\label{eq:calc_average_abs}
\end{eqnarray}
Here, $N$ is the number of realizations and $k$ is the absolute value of ${\bf k}$.

\section{Result}

In this section, we summarize our result for the 21cm bispectrum.

\subsection{Scale-dependence of 21cm bispectrum}

\begin{figure}
   \centering
   \includegraphics[width=1.0\hsize]{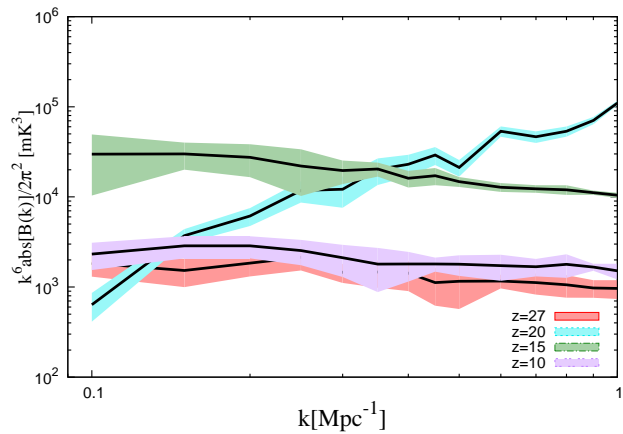}
   \caption{Equilateral type bispectra as functions of wave number at $z =$ 27 (red), 20 (cyan), 15 (green), 10 (purple). The shaded region associated with each line represents 1-$\sigma$ sample variance estimated from 10 realizations.}
\label{fig:bispectrum_abs}
\end{figure}

First, in order to see the scale-dependence of the 21cm bispectrum, we focus on the equilateral shape, that is, $\alpha = 1$ case for the isosceles ansatz discussed in the previous section. We plot the equilateral type bispectrum as a function of wave number $k$ with 1-$\sigma$ sample variance for several redshifts ($z = 10, 15, 20$ and $27$) in Fig.\ref{fig:bispectrum_abs}. Here, we use the normalized bispectrum which is given by $k^6 abs[\overline{B(k)}]$. $z=10$ is a typical redshift during EoR, and $z = 15$ and $20$ are expected to be a transition time from CD to EoR, while $z = 27$ is a typical time during CD. As you can see, the variance is relatively small and the cosmic variance is not so serious for the field size and wavenumbers we chose. From this figure, we can find that, except for the case with $z = 20$, the normalized bispectrum is almost scale-invariant for the equilateral shape. On the other hand, for $z = 20$, the normalised bispectrum has a scale-dependence as $\propto k^2$. Such difference is expected to depend on what component gives a dominant contribution to the 21cm bispectrum.  As we will discuss this issue later, dominant component of the 21cm bispectrum at $z$=20 is matter fluctuations whereas dominant components of the 21cm bispectrum at other redshifts are not matter fluctuations, but spin temperature and neutral fraction fluctuations.  Since the bispectrum of dark matter fluctuations has scale dependence (larger at smaller scales due to the nonlinear gravitational growth), the 21cm bispectrum traces this scale dependence.

\begin{figure*}
   \centering
     \includegraphics[width=1.0\hsize,]{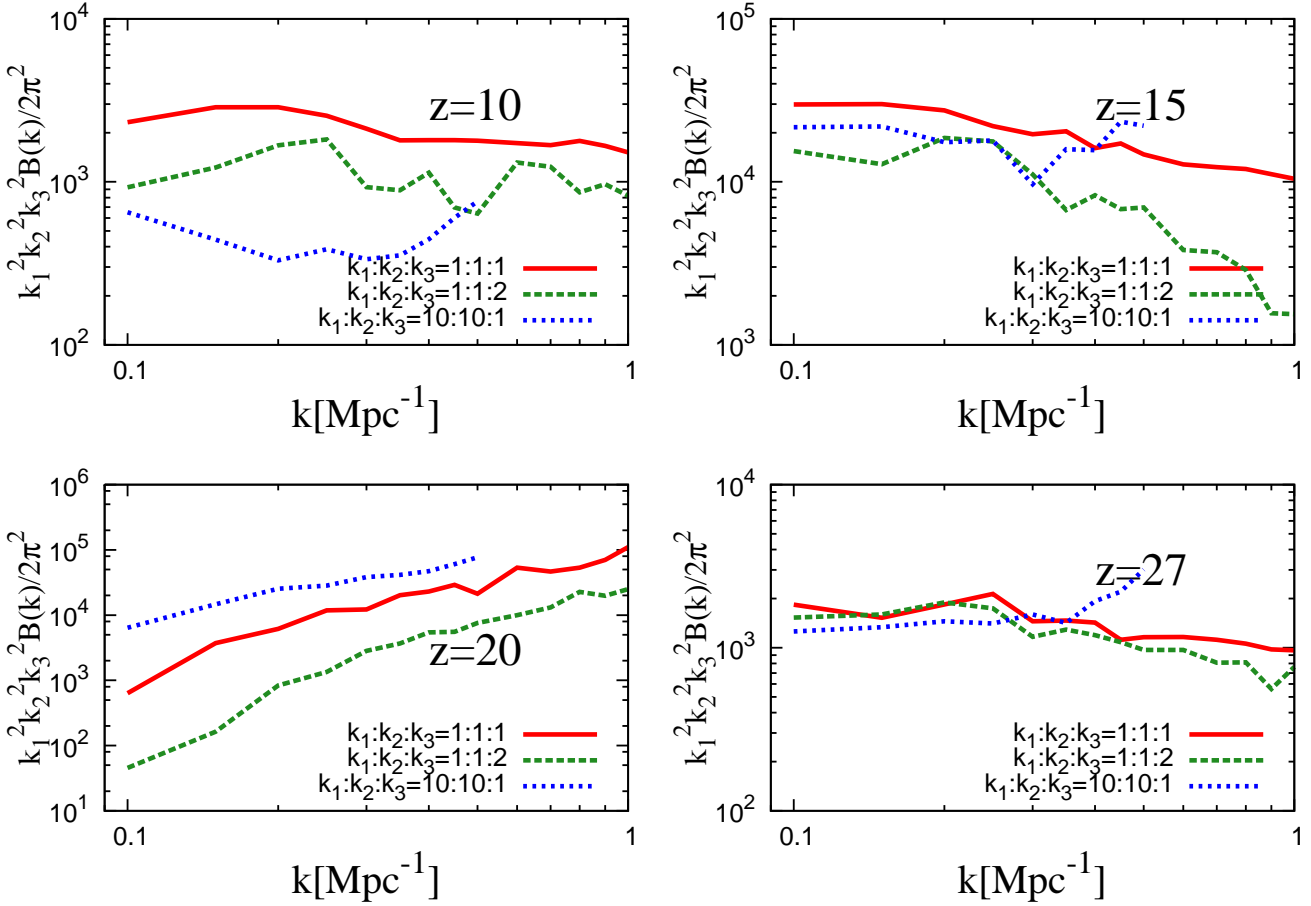}
   \caption{Scale dependence of bispectra for $(k_1:k_2:k_3) = (1:1:1), (1:1:2)$and $(10:10:1)$, as functions of $k_3$.}
\label{fig:bispectrum_squeezed}
\end{figure*}

Next, we show a comparison of bispectra of equilateral ($\alpha = 1$), folded ($\alpha = 1/2$) and squeezed ($\alpha = 10$) types in Fig. \ref{fig:bispectrum_squeezed}. Here, the bispectra are plotted as functions of $k_3$ for several redshifts. From this figure, we can see that the scales and shapes which mostly contribute to the skewness since skewness is the integral of bispectra (see Appendix A). Especially, smaller scales contributes to the skewness at $z=20$ although the bispectrum is nearly scale invariant at other redshifts.  We also find that the squeezed type bispectra at $z$=15, 27 turn upwards at $k>0.3{\rm Mpc}^{-1}$.  The squeezed-type bispectra brings information on smaller scales and matter fluctuations are dominant at smaller scales.  Thus, contribution from matter fluctuations results in these upwards at squeezed type of bispectra. 

\subsection{Redshift evolution of 21cm bispectrum}

\begin{figure}
\centering
\includegraphics[width=1.0\hsize]{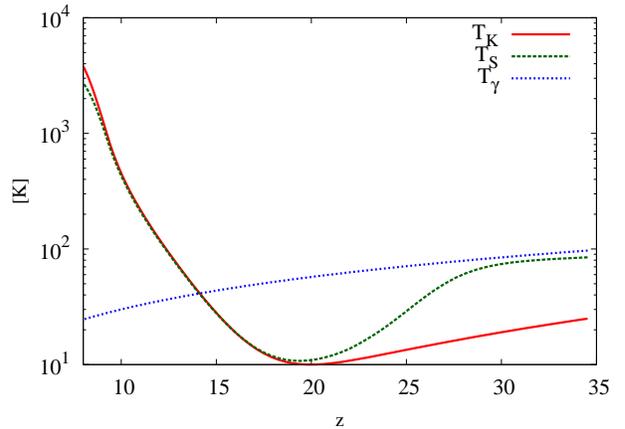}
\caption{The redshift evolution of temperatures. We show evolution of the kinetic temperature(red), the spin temperature(green) and the CMB temperature(blue).}
\label{fig:xH_Ts}
\end{figure}

\begin{figure}
   \centering
   \includegraphics[width=0.6\hsize]{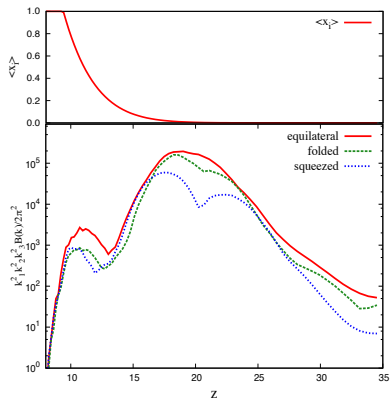}
   \caption{({\it Top}) ionization history in our model.  ({\it Bottom}) Comparison of bispectra of typical triangle configurations. We fix $k = 1.0~{\rm Mpc}^{-1}$ and take $\alpha = 1$ (equilateral: red solid line), $\alpha = 1/2$ (folded: green dashed line) and $\alpha = 10$ (squeezed: blue dotted line) for the isosceles ansatz.}
\label{fig:bispectrum_triangle3}
\end{figure}

Next, we consider redshift evolution of 21cm bispectrum. Before we show redshift evolution of the 21cm bispectrum, as reference, we show redshift evolution of temperatures in our model in Fig.\ref{fig:xH_Ts}.  We also show ionized evolution and the bispectra as functions of redshift for several $\alpha$ in Fig. \ref{fig:bispectrum_triangle3}: the equilateral shape ($\alpha = 1$), the folded shape ($\alpha = 1/2$) and the squeezed shape ($\alpha = 10$) with $k = 1.0~{\rm Mpc}^{-1}$. For the equilateral and folded cases, we can see two peaks located at around $z = 20$ and $12$. These peaks can also be seen in the power spectrum of the brightness temperature fluctuations, $P_{21}(k)$, with $k \simeq 1.0~{\rm Mpc}^{-1}$ (see, e.g., our previous paper \cite{2015MNRAS.451.4986S}). On the other hand, in case with the squeezed shape, three peaks appear at around $z = 23$, $17$, and $12$. This feature is similar to that of the power spectrum with $k \simeq 0.1~{\rm Mpc}^{-1}$ \cite{2015MNRAS.451.4986S}. For the squeezed type, we take the parameter $\alpha$ to be $10$ and this means $k_3 = 0.1~{\rm Mpc}^{-1}$. Hence, the squeezed-type 21cm bispectrum is expected to be described in terms of not only the power spectrum with larger two wave number ($k_1$ and $k_2$ in our case) but also that with smaller one wave number ($k_3$ in our case) and also it would have the information about the correlation between the long and short wavelength modes in Fourier space or local non-linearity in real space.  The dip at $z\sim 20$ for the $\alpha=10$ case results in mode coupling between long and short wavelengths.  As we show later in Fig.\ref{fig:bispectrum_component_equilateral}, the 21cm bispectrum of equilateral type at large scale($k=0.1{\rm Mpc}^{-1}$) also shows a dip at $z\sim 20$ as similar as the power spectrum as function of redshift and the $\alpha=10$ traces this feature.  Therefore, we can conclude that the squeezed type bispectrum has information both on large scale and small scale. 

We will also investigate what physics cause such a correlation between the long and short wavelength modes in the 21cm bispectrum in later subsection.

\subsection{Decomposition of 21cm bispectrum}
\label{sec:component}

As we saw in Eq. (\ref{eq:brightness}), the fluctuations in the brightness temperature are contributed not only from the matte density field, but also from the fluctuations of the spin temperature and neutral fraction, aside from the gradient of peculiar velocity which we neglect here. In this section, we decompose the bispectrum into the contributions from these components. 

We can rewrite Eq. (\ref{eq:brightness}) as,
\begin{equation}
\delta T_b({\bf x})
= \overline{\delta T}_b (1+\delta_{x_{\rm H}}({\bf x}))
  (1 + \delta_m({\bf x})) (1 + \delta_{\eta}({\bf x})),
\label{eq:brightness_component}
\end{equation}
where $\overline{\delta T_b}$ is the average brightness temperature and evaluated as,
\begin{equation}
\overline{\delta T_b}
= 27 \overline{x_{\rm H}} \overline{\eta}
  \left( \frac{1+z}{10} \right)^{1/2}
  \left( \frac{0.15}{\Omega_m h^2} \right)^{1/2}
  \left( \frac{\Omega_b h^2}{0.023} \right),
\end{equation}
with $\overline{x_{\rm H}}$ being the volume average of $x_{\rm H}$. Here we characterize the contribution of the spin temperature $T_s$ by a new variable $\eta = 1 - T_\gamma/T_{\rm S}$ \cite{2015MNRAS.451.4986S}. By using this parameter, we can take into account a non-linear relation between the spin and brightness temperatures linearly. The volume average of $\eta$ is represented by $\overline{\eta}$. Note that when $\delta_{T_{\rm S}} \ll 1$, we have,
\begin{eqnarray}
\delta_{\eta} \simeq
\frac{T_\gamma / \bar{T}_{\rm S}}{1 - T_\gamma / \bar{T}_{\rm S}}
\delta_{T_{\rm S}}.
\end{eqnarray}

Using Eq. (\ref{eq:brightness_component}), we can decompose the brightness temperature bispectrum into auto- and cross-correlation of $\delta_m, \delta_{x_{\rm H}}$ and $\delta_\eta$:
\begin{eqnarray}
B_{\delta T_b}
&=& (\overline{\delta T}_{b})^{3}
    [B_{\delta_m \delta_m \delta_m}
     + B_{\delta_{x_{\rm H}} \delta_{x_{\rm H}} \delta_{x_{\rm H}}}
     + B_{\delta_{\eta} \delta_{\eta} \delta_{\eta}} \nonumber \\
& &  + ({\rm cross ~ correlation ~ terms}) \nonumber \\
& &  + ({\rm higher ~ order ~ terms})].
\label{eq:bispectrum_component}
\end{eqnarray}
In the above equation, the cross correlation terms and the higher order terms come from the fact that the brightness temperature is expressed as Eq. (\ref{eq:brightness_component}) and they should appear even if the statistics of $\delta_{m}, \delta_{x_{\rm H}}$ and $\delta_\eta$ are completely Gaussian. In this sense, the first three terms in the above expression, which are the auto-bispectra of $\delta_m, \delta_{x_{\rm H}}$ and $\delta_\eta$, should be corresponding to the intrinsic non-Gaussian features of these components and we focus on these auto-bispectra below.

\begin{figure*}
   \centering
   \includegraphics[width=1.0\hsize]{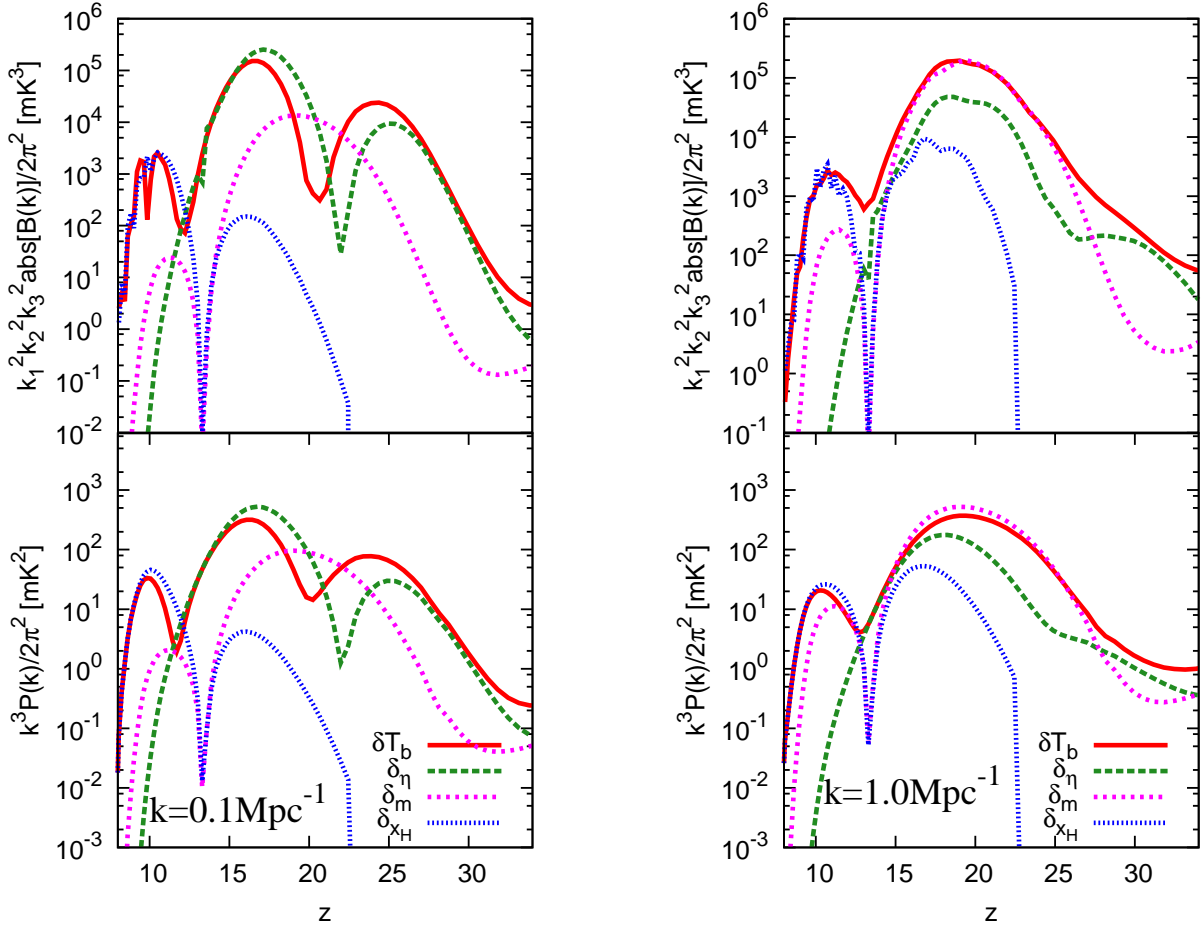}
   \caption{Components of 21cm bispectrum for equilateral type: the brightness temperature (red), the contribution from eta (green), matter fluctuations (blue) and neutral hydrogen fraction (magenda). }
\label{fig:bispectrum_component_equilateral}
\end{figure*}

In Fig. \ref{fig:bispectrum_component_equilateral}, we plot the brightness temperature bispectrum and the above auto-bispectra terms for equilateral type as functions of redshift (upper panels). From this figure, we can see that the total bispectra are mostly contributed from the auto-bispectra of the matter density field, the fluctuations of the spin temperature and the neutral fraction, which are expressed as the first three terms in Eq. (\ref{eq:bispectrum_component}), for all redshifts. For comparison, the evolution of power spectra is also shown (lower panels) and we find that the behavior of each component is very similar between bispectrum and power spectrum. Such a correspondence is highly non trivial, since the bispectrum and power spectrum reflect different aspects of the statistical properties of the fluctuations as we have mentioned.

Let us try to interpret the behavior of bispectra, comparing that of power spectra which was detailed in our previous work \cite{2015MNRAS.451.4986S}. First, fluctuations in neutral hydrogen fraction appear when reionization begins and become dominant as reionization proceeds ($z \lesssim 12$). The dip at $z \sim 14$ corresponds to the redshift when the average spin temperature becomes equal to the CMB temperature and the average brightness temperature $\overline{\delta T_{b}}$ vanishes. This dip appears in the contribution of matter fluctuations for the same reason. Thus, this dip is independent of the properties of fluctuations and this is why both the power spectra and bispectra from $\delta_{x_{\rm H}}$ and $\delta_m$ have a dip at the same redshift. Note that we can also see the dip at $z\sim 8$ in the 21cm bispectrum at $k=0.1{\rm Mpc}^{-1}$.  It would be due to sample variance coming from calculation of ionized bubbles because evolutions of ionized bubbles are different among other realizations.

On the other hand, the spin temperature fluctuations are negligible at low redshifts ($z \lesssim 10$), because spin temperature is much higher than CMB temperature everywhere, that is, $\eta = 1 - T_{\gamma}/T_{\rm S}$ is very close to unity and independent of $T_{\rm S}$. However they substantially contribute at higher redshifts ($z \gtrsim 14$) and have two peaks at $z \sim 15$ and $z \sim 25$ at large scales while the higher-redshift peak is much less noticeable at small scales. The dip at $z \sim 23$ is induced by the onset of X-ray heating. At higher redshifts ($z \gtrsim 23$), the spin temperature in dense region is lower than the average due to the WF effect, which couples the spin temperature to the kinetic temperature which is much lower than the CMB temperature. As a consequence, the probability distribution function (pdf) of spin temperature is negatively skewed at this epoch. Contrastingly, at lower redshifts ($z \lesssim 23$), X-ray heating becomes effective for our parameter set and the spin temperature in dense region rises rapidly so that the skewness of the pdf changes its sign. Thus, at the onset of X-ray heating ($z \sim 23$), the pdf becomes close to Gaussian with relatively small width, that is, the the skewness vanishes and the variance has a local minimum. Thus, in \cite{2015MNRAS.451.4986S}, it was suggested that the sign of skewness can be an indicator of the effectiveness of X-ray heating. The skewness and the variance have minima that coincide in redshift, similarly, it is natural that the dips in the bispectrum and the power spectrum are coincident with each other.

Before EoR, the bispectrum, as well as the power spectrum, is mostly dominated by spin temperature fluctuations at large scales and will be a good probe of astrophysical effects such as the WF effect and X-ray heating. On the other hand, at small scales, the matter fluctuations are dominant and the bispectrum is of cosmological interest because it is induced by gravitational nonlinearity and, possibly, primordial non-Gaussianity (see, e.g., \cite{Scoccimarro:2000sn,Bernardeau:2001qr,2009ApJ...703.1230J}).

\begin{figure*}
\centering
\includegraphics[width=0.6\hsize]{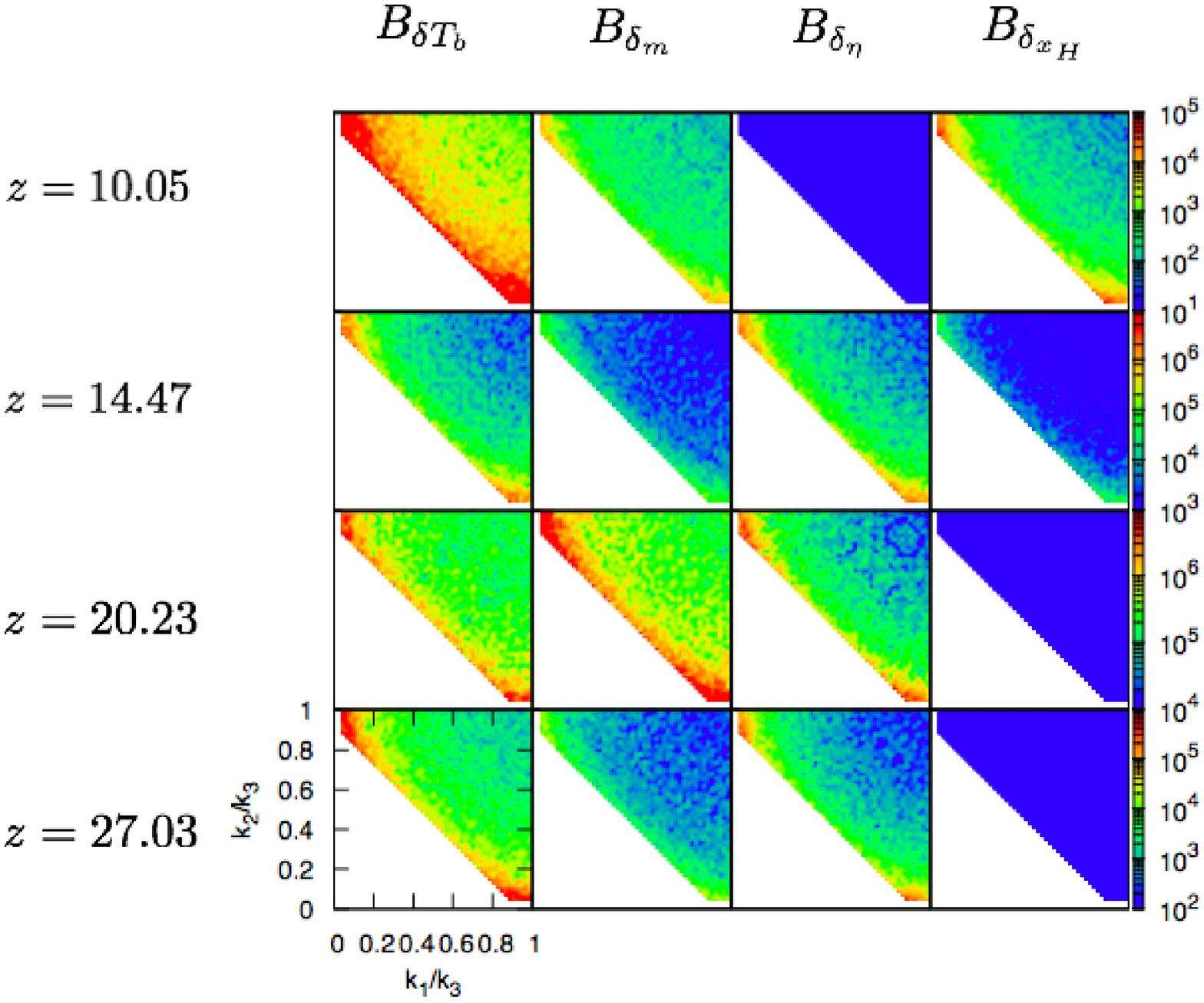}
\includegraphics[width=0.6\hsize]{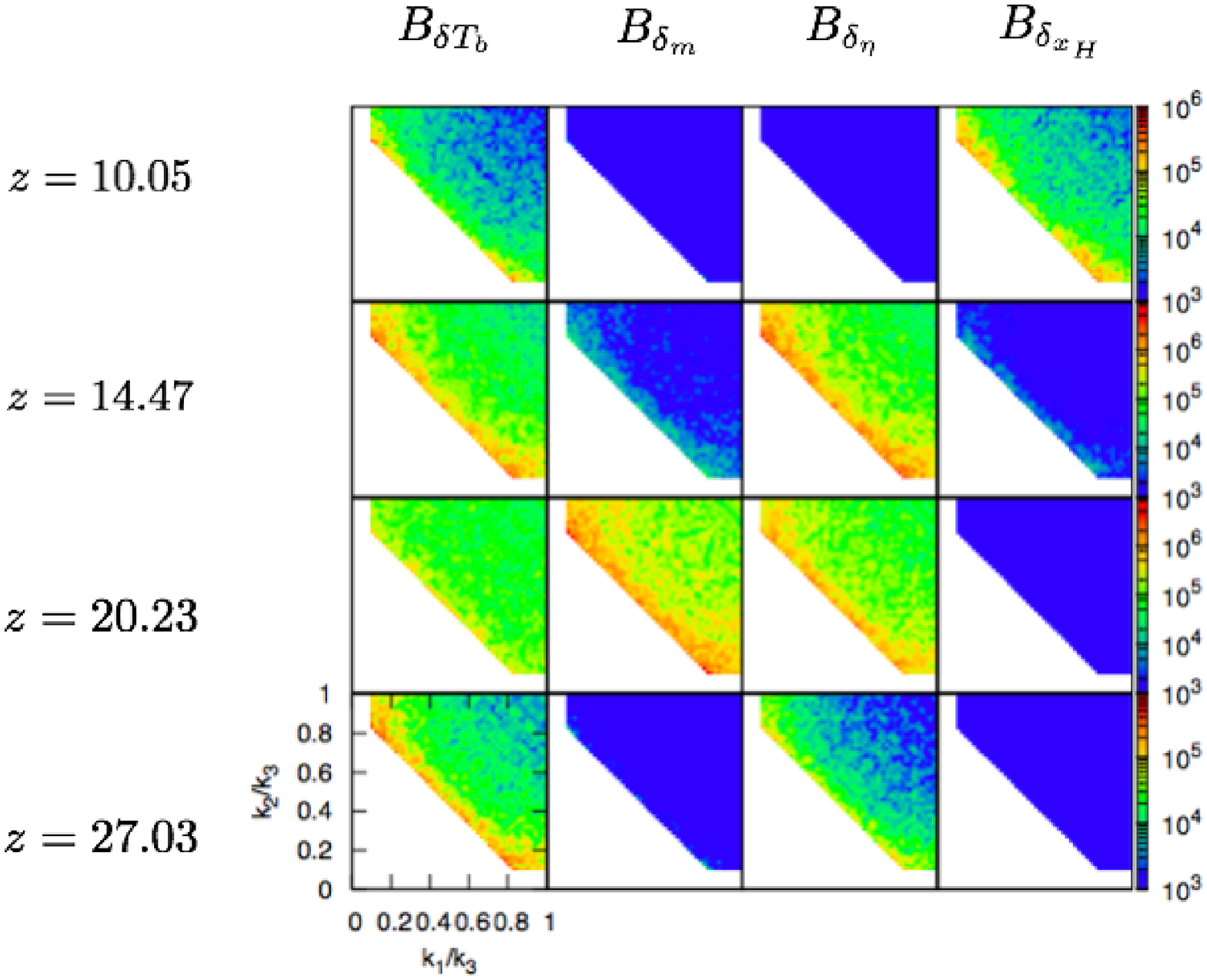}
\caption{(Top) Contours of the total bispectrum and its components in $k_1/k_3$-$k_2/k_3$ plane with $k_3 = 1.0~{\rm Mpc^{-1}}$. (Bottom) Contours of the total bispectrum and its components in $k_1/k_3$-$k_2/k_3$ plane with $k_3 = 0.4~{\rm Mpc^{-1}}$.}
\label{fig:bispectrum_component_free}
\end{figure*}

Next, we focus on the shape dependence of the total bispectrum and its components. Fixing $k_3 = 1.0~{\rm Mpc}^{-1}$, we plot contours of the bispectra in $(k_1/k_3)$-$(k_2/k_3)$ plain in top of Fig. \ref{fig:bispectrum_component_free}. Note that we do not use the normalised bispectrum, $k_1^2 k_2^2 k_3^2 B(k_1,k_2,k_3)$, but the unnormalized bispectrum, $B(k_1,k_2,k_3)$, here. We can see in what configuration of triangle the bispectra are strong. Here it should be noted that the triangle condition is not satisfied in the blank region and that the contours are symmetric with respect to a line $k_1/k_3 = k_2/k_3$.

At $z=10.05$ when EoR has proceeded to some extent($x_{i}=0.77$), the total bispectrum is strong at folded and squeezed types. The contribution from neutral hydrogen fraction fluctuations is dominant at these configurations, while matter fluctuation is dominant at equilateral type. At $z=14.47$, the dominant contribution comes from the spin temperature fluctuations and it is largest at squeezed type. The situation is similar at $z=27.03$. At $z=20.23$, both squeezed and folded type of the total bispectrum are strong. The contributions from matter and spin temperature fluctuations are comparable at these configurations, while the former is dominant at equilateral type.


We also show the contour for $k_3 = 0.4~{\rm Mpc^{-1}}$ in bottom of Fig.\ref{fig:bispectrum_component_free}. Compared with the case of $k_3 = 1.0~{\rm Mpc^{-1}}$, contributions from both matter and spin temperature fluctuations are significant at $z=20.23$. On the other hand, we find that the contribution from fluctuations of neutral hydrogen fraction at $z=10.05$ is clear compared with the case of $k_3 = 1.0~{\rm Mpc^{-1}}$. This helps us to subtract the information on neutral hydrogen and it is better to see larger scales if we would like to know the information on neutral hydrogen fluctuations.

\section{Discussion \& Summary}

In this paper, we investigated the 21cm bispectrum as a method to measure non-Gaussianity of brightness temperature field.

First, we have shown the scale-dependence of the 21cm bispectrum for the ``equilateral-shape" at some redshifts. We found that the normalised 21cm bispectrum seems not to have any characteristic scale in $0.1 \lesssim k / {\rm Mpc}^{-1} \lesssim 1.0$ for each redshift. For $z = 10, 15, 27$, the normalised bispectrum is almost scale-invariant, while for $z = 20$ it has a scale-dependence as $k^6 B \propto k^2$.

We have also shown the redshift evolution of the 21cm bispectrum with fixed $k$ for three types of the shape in $k$-space. We found that the redshift evolution of the 21cm bispectrum for the equilateral and folded shapes basically traces that of the 21cm power spectrum, but in case with the squeezed shape, we could see a different behavior and it can be understood by considering the coupling between the large- and small-scale modes.

Then, we studied the 21cm bispectra by decomposing it into the contributions from the matter density field, the fluctuations in the spin temperature and the neutral fraction. From the redshift evolution, we found the dominant component at each redshift and scale. We also show the shape dependence of each component and compared it with that of total 21cm bispectrum. The shape dependence of each component looks similar to each other, but a slight difference also exists. Hence, by future precise observation it is expected that we would obtain the information about the non-Gaussian feature of these components separately.

As far as the matter bispectrum is concerned, there have been a lot of works which discuss the shape-dependence by using the second order perturbation theory and also numerical N-body simulation. Although most of the works focus on the matter bispectrum at the lower redshift ($z \simeq 1.0$) or higher redshift ($30 \lesssim z \lesssim 100$ in the dark age) (e.g., \cite{Scoccimarro:2000sn,Bernardeau:2001qr,2009ApJ...703.1230J,Sefusatti:2010ee,Cooray:2006km,2015arXiv150604152M}), we can find that the shape of the matter bispectrum in our result is basically consistent with these previous works by extrapolation. By using the second order perturbation theory, the matter bispectrum can be expressed as $B_m (k_1, k_2, k_3) \propto P_m (k_1) P_m (k_2) + 2~{\rm perms.}$ with $P_m(k)$ being the matter power spectrum. For $0.1 ~ {\rm Mpc}^{-1} \lesssim k$, the matter power spectrum behaves as $\propto k^{-2 \sim 3}$. Based on this fact and the isosceles ansatz ($k_1 = k_2 = k = \alpha k_3 $), we have $B_m (k, \alpha) \propto (1 + 2 \alpha^{2 \sim 3}) k^{-4 \sim 6}$, and hence the unnormalised matter bispectrum becomes larger as $\alpha$ increases is the largest in the squeezed shape \cite{2009ApJ...703.1230J}.

Based on the above discussion about the matter bispectrum, let us revisit the behavior of the 21cm bispectrum as shown in Fig. \ref{fig:bispectrum_abs}. As we have mentioned, in contrast to other redshifts, for $z = 20$ the normalised 21cm bispectrum has a scale-dependence as $k^2$ in the equilateral shape. From Fig. \ref{fig:bispectrum_component_equilateral}, we find that at $z = 20$ the matter contribution relatively dominates over the 21cm bispectrum, and hence the behavior of the 21cm bispectrum is expected to trace that of the matter bispectrum at this redshift. Based on the expression obtained from the second order perturbation theory, the scale-dependence of the normalised matter bispectrum can be estimated as $\propto k^6 \times k^{-4 \sim 6} = k^{0 \sim 2}$. Hence, the behavior of the matter bispectrum could explain that of the 21cm bispectrum at $z = 20$.

Note that the Zel'dovich approximation is used to solve density evolution in 21cmFAST.  This approximation is imperfect to calculate the bispectrum and it is desirable to use N-body simulations to estimate the bispectrum accurately.  However, at scales less than $k \sim 5 {\rm Mpc}^{-1}$ at $z=7-20$ , density evolution with the Zel'dovich approximation coincides with that obtained by N-body simulations\cite{2011MNRAS.411..955M}.

Naively, the spatial distributions of the spin temperature and neutral fraction should be considered as tracers of matter density field, that is, they could be treated equally with those of the halos and galaxies. Based on this consideration, the simplest way to express the spatial distributions of the spin temperature and neutral fraction is introducing a bias parameter, such as $\delta_i \propto b_i \delta_m$ ($i = \eta$ and $x_{\rm H}$). If such a bias parameter is scale-independent, the behaviors of the bispectra of the spin temperature and neutral fraction are completely the same as that of the matter density field. However, we can see slight differences between these components in Fig. \ref{fig:bispectrum_component_free} and also \ref{fig:bispectrum_abs}. Hence, we expect that the bias parameter should have non-trivial scale-dependence due to the non-linear or non-local transfer from the matter density field to the spin temperature and neutral fraction. We need to investigate this issue more deeply in future work.

\begin{figure*}
   \centering
   \includegraphics[width=1.0\hsize, bb=0 0 380 290]{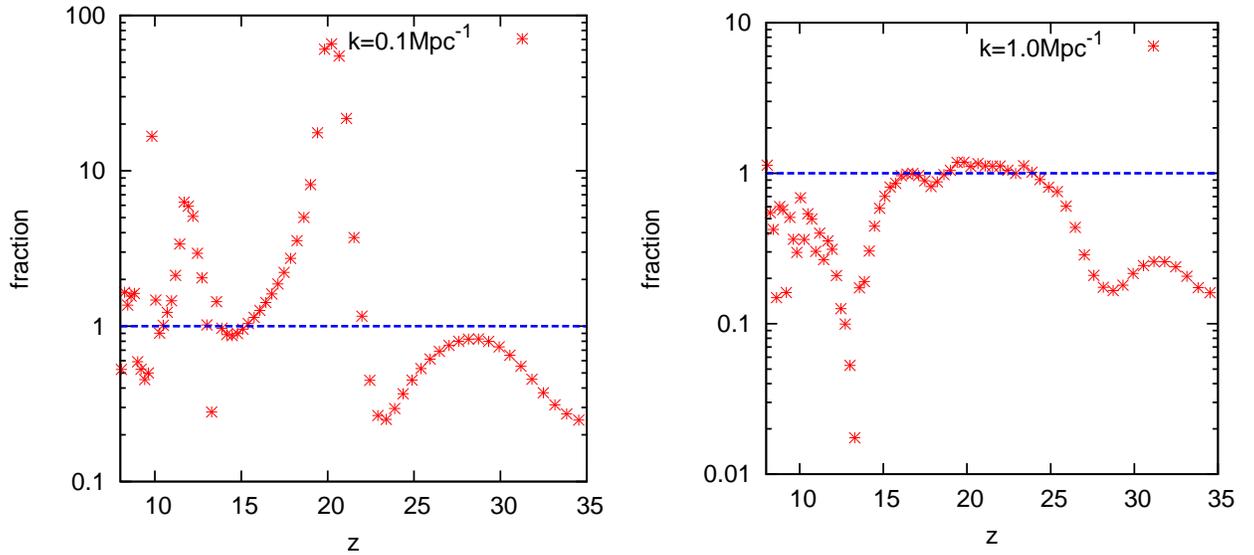}
   \caption{ This shows the fractional difference between sum of three component terms and total 21cm bispectrum at $k=0.1{\rm Mpc^{-1}}(left)$ and $k=1.0{\rm Mpc^{-1}}(right)$. Here, we plot the $(B_{\delta_{m}}+B_{\delta_{\eta}}+B_{x_{H}})/B_{\delta T_{b}}$}
\label{fig:fraction}
\end{figure*}

Here, we briefly discuss the cross correlation higher order terms in eq.(\ref{eq:bispectrum_component}).  Although we only plot auto correlation term in Fig.\ref{fig:bispectrum_component_equilateral}, neglected cross correlation and higher order terms are of the same order as the auto correlation terms.  We show the ratio between the sum of three component terms and total 21cm bispectrum as function of redshift in Fig.\ref{fig:fraction}.  As you can see from this figure, this ratio is not unity since we do not include the cross correlation and higher order terms, and hence, in practice, these terms can not be neglected. However, in our study, we focus on only auto correlation terms in order to study contribution from each component.  Although we expect that what cross correlation terms are effective at each epoch from Fig.\ref{fig:bispectrum_component_free} (e.g. cross correlation between fluctuations neutral hydrogen fraction and matter density might become effective at the EoR ($z\sim 10$)), detailed study is our future work.

The detectability of bispectrum is of critical interest. In our previous work \cite{2015MNRAS.451.4785Y}, we estimated the signal-to-noise ratio of bispectrum, developing a formalism to calculate the bispectrum contributed from thermal noise.  We find that the SKA1 has enough sensitivity for both equilateral and isosceles($K = |k_1| = |k_2|, k = |k_3|$ with K=0.06 ${\rm Mpc}^{-1}$) triangles for $k\lesssim 0.3$ at $z$=8-17, while LOFAR will have sensitivity for the peaks of the bispectrum as a function of redshift. Actually, galactic and extragalactic foreground will be a serious obstacle just as in the case of power spectrum and should be studied in detail.

There are some other approaches to measure non-Gaussianities in the brightness temperature field. For example, some studies focus on topological structure of brightness temperature field such as Minkowski functionals \cite{2006MNRAS.370.1329G,2008ApJ...675....8L,2014JKAS...47...49H,yoshi2015}. This method is complementary to higher order statistics.

\section*{Appendix A}

In this section, we derive the relation between bispectrum and skewness. When we define the fluctuation of brightness temperature $\delta T_{b}$  as $\delta=(\delta T_{b}-\overline{\delta T_b})/\overline{\delta T_b}$, the skewness of brightness temperature is defined by 
\begin{eqnarray}
\gamma&=&\frac{1}{N}\sum_{i=1}^{N}(\delta T_{b,i}-\overline{\delta T_{b}})^{3}\\
&=&\frac{(\overline{\delta T_{b}})^{3}}{N}\sum\delta^{3} \nonumber \\
&=&(\overline{\delta T_{b}})^{3}\langle \delta^{3}\rangle. \nonumber
\label{eq:skewness1}
\end{eqnarray}
Here $N$ is the number of pixel. The definition of three-point correlation function for brightness temperature $\xi$ is expressed by $\xi({\bf r_{1}},{\bf r_{2}})=\langle \delta({\bf x})\delta({\bf x}+{\bf r_{1}}) \delta({\bf x}+{\bf r_{2}})\rangle$.  We can connect this three-point correlation function with ensemble average of $\delta^{3}$ such as $\langle \delta^{3} \rangle =\xi(0,0)$ .  We know the relation between correlation function and bispectrum from Wiener-Khintchine relation described by
\begin{equation}
\xi({\bf r_{1}},{\bf r_{2}})=\int \frac{d^{3}k_{1}}{(2\pi)^{3}}\int \frac{d^{3}k_{2}}{(2\pi)^{3}}e^{i{(\bold k_{1}}\cdot{\bf r_{1}}+{\bf k_{2}}\cdot{\bf r_{2}})}B({\bf k_{1}},{\bf k_{2}},-{\bf k_{1}}-{\bf k_{2}}).
\label{eq:WK}
\end{equation}
Therefore, $\xi(0,0)$ can be expressed with bispectrum by 
\begin{equation}
\xi(0,0)=\int \frac{d^{3}k_{1}}{(2\pi)^{3}}\int \frac{d^{3}k_{2}}{(2\pi)^{3}}B({\bf k_{1}},{\bf k_{2}},-{\bf k_{1}}-{\bf k_{2}}).
\label{eq:WK2}
\end{equation}
By using equation(\ref{eq:WK}) and (\ref{eq:WK2}),  we can derive the relation between skewness and bispectrum as,
\begin{equation}
\gamma=(\overline{\delta T_{b}})^{3}\int \frac{d^{3}k_{1}}{(2\pi)^{3}}\int \frac{d^{3}k_{2}}{(2\pi)^{3}}B({\bf k_{1}},{\bf k_{2}},-{\bf k_{1}}-{\bf k_{2}}).
\label{eq:skewness_bispectrum}
\end{equation}

\section*{Acknowledgement}
We would like to thank K. Hasegawa and D. Nitta for useful comments. This work is supported by Grant-in-Aid from the Ministry of Education, Culture, Sports, Science and Technology (MEXT) of Japan, Nos. 24340048(K.T. and K.I.) and 26610048. (K.T.), No. 25-3015(H.S.) and 15K17659 (S.Y.).



\label{lastpage}

\end{document}